\documentstyle[epsfig]{ioplppt}

\def\lapproxeq{\lower .7ex\hbox{$\;\stackrel{\textstyle                                         
<}{\sim}\;$}}                                         
\def\gapproxeq{\lower .7ex\hbox{$\;\stackrel{\textstyle                                         
>}{\sim}\;$}}                                         
       
\newcommand{\be}{\begin{equation}}       
\newcommand{\ee}{\end{equation}}       
\newcommand{\bea}{\begin{eqnarray}}       
\newcommand{\eea}{\end{eqnarray}}

\begin{document}
\titlepage
\begin{flushright}
DTP/99/108 \\
December 1999 \\
\end{flushright}

\begin{center}
\vspace*{2cm}
{\Large \bf The $R$ ratio in $e^+ e^-$, the determination of $\alpha (M_Z^2)$ and a possible 
non-perturbative gluonic contribution} \footnote{Contribution to the Proceedings of the UK Phenomenology Workshop on Collider Physics, Durham,
UK, 19-24 September 1999, to be published in J. Phys. G.}\\

\vspace*{1cm}
A.D. Martin, J. Outhwaite and M.G. Ryskin \\

\vspace*{0.5cm}

Department of Physics, University of Durham, Durham, DH1 3LE, UK.

\end{center}

\vspace*{0.3cm}

\begin{abstract}
We review the determination of the QED coupling at the $Z$ pole, which is a crucial parameter for electroweak 
theory.  We include recent $e^+ e^- \rightarrow$ hadron data from Novosibirsk and Beijing to re-evaluate $\alpha (M_Z^2)$.  
We find $\alpha (M_Z^2)^{-1} = 128.973 \pm 0.035$ or $128.934 \pm 0.040$ according, respectively, to whether 
inclusive or exclusive $e^+ e^- \! \rightarrow$ hadron data are used respectively in the interval $1.4 \lapproxeq 
\sqrt{s} \lapproxeq 2.1$~GeV.  The error is mainly due to uncertainties in the data in the low energy region, 
$\sqrt{s} \lapproxeq 2.5$~GeV.  We find that no advantage is obtained by analytic continuation of the dispersion relation 
into the complex $s$ plane.  We show that the hints of structure for $\sqrt{s} \sim 2.5$~GeV may be evidence of a 
non-perturbative gluonic contribution to $R(s)$.
\end{abstract}

The value of the QED coupling at the $Z$ pole, $\alpha(M_Z^2)$, is the poorest known of the three 
parameters ($G_F, M_Z, \alpha (M_Z^2)$) which define the standard electroweak model.  Indeed it is the 
precision to which we know $\alpha (M_Z^2)$ which limits the accuracy of the indirect prediction of the 
mass $M_H$ of the (Standard Model) Higgs boson \cite{Z}.

The value of $\alpha (M_Z)$ is obtained from
\be
\label{eq:a1}
\alpha^{-1} \; \equiv \; \alpha (0)^{-1} \; = \; 137.0359895 (61)
\ee
using the relation
\be
\label{eq:a2}
\alpha (s)^{-1} \; = \; \left ( 1 - \Delta \alpha_{\rm lep} (s) - \Delta \alpha_{\rm had}^{(5)} (s) - 
\Delta \alpha^{\rm top} (s) \right ) \: \alpha^{-1},
\ee
where the leptonic contribution to the running of the $\alpha$ is known to 3 loops \cite{S}
\be
\label{eq:a3}
\Delta \alpha_{\rm lep} (M_Z^2) \; = \; 314.98 \times 10^{-4}.
\ee
From now on we omit the superscript (5) on $\Delta \alpha_{\rm had}$ and assume that it corresponds to five flavours.  
We will include the contribution of the sixth flavour, 
$\Delta \alpha^{\rm top} (M_Z^2) = -0.76 \times 10^{-4}$, at the end.  To determine the hadronic contribution 
we need to evaluate
\be
\label{eq:a4}
\Delta \alpha_{\rm had} (s) \; = \; - \frac{\alpha s}{3 \pi} \: P \int_{4m_\pi^2}^{\infty} \: 
\frac{R (s^\prime) ds^\prime}{s^\prime (s^\prime - s)}
\ee
at $s = M_Z^2$, where $R = \sigma (e^+ e^- \rightarrow {\rm hadrons})/\sigma (e^+ e^- \rightarrow \mu^+ \mu^-)$.

The early determinations \cite{EJ} of $\Delta \alpha_{\rm had} (M_Z^2)$ made maximum use of the $e^+ e^-$ measurements 
of $R(s)$, using the sum of the exclusive hadronic channels $(e^+ e^- \rightarrow 2\pi, 3\pi, \ldots K\bar{K}, \ldots)$ 
for $\sqrt{s} \lapproxeq 1.5$~GeV and the inclusive measurement of $\sigma (e^+ e^- \rightarrow {\rm hadrons})$ at larger 
energies.  However, in the last few years the determinations of $\Delta \alpha_{\rm had}$ from (\ref{eq:a4}) have relied more and 
more on theoretical input.  First perturbative QCD was used to better describe $R(s)$ (for $\sqrt{s} > 3$~GeV) in energy 
regions above the resonances to the next flavour threshold \cite{MZ}.  Then, encouraged by the success of perturbative QCD to 
describe $\tau$ decay, it was used down to 1.8~GeV, across a region with sparse data on $R(s)$, giving \cite{KS}
\be
\label{eq:a5}
\Delta \alpha_{\rm had} (M_Z^2) \; = \; (277.5 \pm 1.7) \times 10^{-4},
\ee
where $\pm 10^{-4}$ comes from the uncertainty in perturbative QCD.

Recent methods to calculate $\alpha (M_Z^2)$ have used the analytic behaviour of $\Pi (s)$ in the complex $s$ plane.  
$\Pi$ is the hadronic contribution to the photon vacuum polarisation amplitude (or two-point correlator), where
\be
\label{eq:a6}
\Delta \alpha_{\rm had} (s) \; = \; - 4 \pi \alpha {\rm Re} \Pi (s), \quad\quad R(s) \; = \; 12 \pi {\rm Im} \Pi (s).
\ee
Groote et al. \cite{GRO} introduce special polynomial weight functions, 
which are arranged to suppress data contributions in specific regions and to essentially replace them by perturbative 
QCD (together with non-perturbative gluon and light quark condensate contributions).  For example, for a specific 
interval $(s_a, s_b)$, the idea is to deform the contour of integration so that the QCD contribution is evaluated 
around circular contours of radii $s_a$ and $s_b$, away from singularities on the $s_a$, $s_b$ cut\footnote{The QCD 
contribution was also evaluated around a circular contour in the $s$ plane in Ref.~\cite{ERLER}.}.  The accuracy 
of this approach was improved in \cite{DH} to give an error on $\Delta \alpha_{\rm had} (M_Z^2)$ of $\pm 1.6 \times 
10^{-4}$, close to the error of the straightforward estimate of (\ref{eq:a5}).  

We will see below that QCD does not reproduce the structure of the data for $R(s)$ in the $1.8 < \sqrt{s} < 3$~GeV region.  
This region itself gives at least an error of $\pm 10^{-4}$, which is additional to the errors of \cite{KS,DH}.  
If the QCD description fails to reproduce the data, in an average sense in this interval, then nothing is gained by 
deforming the contour of integration\footnote{Due to Cauchy's theorem the result is independent of the choice of contour.}
using the trick of \cite{GRO,DH}.  It is just not possible to circumvent these measurements of $R(s)$, and their uncertainties, 
in this way.

In an interesting development Jegerlehner et al. \cite{EJKV,J} calculate the value of $\Delta \alpha_{\rm had}$ at 
the large negative scale $s = - M_Z^2$ and then analytically continue around the {\it large} circle to positive 
$s = M_Z^2$.  The QCD error comes mainly from the uncertainty in $\alpha_S (M_Z^2)$, and the neglect of $\alpha_S^4$ 
and higher order corrections.  The extra error due to the analytic continuation around the {\it large} circle of 
radius $\simeq M_Z$ is very small.  At first sight it appears that we have dispensed with the data for $R(s)$ 
altogether, and hence determine $\Delta \alpha_{\rm had} (M_Z^2)$ with an error $\lapproxeq 10^{-4}$ arising just from 
QCD.

Unfortunately there are sizeable errors in $\Pi (-s)$ due to the uncertainty in what to take for the light quark 
masses; even the uncertainty in the charm threshold, $4m_c^2$, gives a sizeable error.  Moreover the constant term 
in the $\alpha_S^3$ contribution is not known yet.  However these problems disappear for the derivative, the 
Adler function,
\be
\label{eq:a7}
D (-s) \; = \; - 12 \pi^2 s \: \frac{d \Pi (-s)}{ds} 
\ee
or for the discontinuity
\be
\label{eq:a8}
R (s) \; = \; \frac{6\pi}{i} \: {\rm disc} \Pi (s) \; = \; 12 {\rm Im} \Pi (s).
\ee
Indeed this was recognized in \cite{EJKV} where $D$ was evaluated for negative $s$ using perturbative QCD (and parton 
condensate contributions).  These authors checked that the theoretical values for $D(-s)$ in the space-like region agree 
well with the \lq\lq direct\rq\rq~evaluation in which the data for $R(s)$ are used to 
determine $\Delta \alpha_{\rm had} (-s)$.  The agreement was found to be good even for $\sqrt{s}$ as low as 2~GeV.  
Unfortunately this is for the derivative, and not the value, of $\Delta \alpha_{\rm had}$.  To determine $\alpha (M_Z^2)$ 
Jergerlehner therefore evaluates
\be
\label{eq:a9}
\fl \Delta \alpha_{\rm had} (-M_Z^2) \; = \; \left [\Delta \alpha_{\rm had} (-M_Z^2) - \Delta \alpha_{\rm had} (-s_0) \right 
]^{\rm QCD} \: + \: \Delta \alpha_{\rm had} (-s_0)^{\rm data}
\ee
where the QCD contribution is accurately known in terms of $D$ and so the error in $\alpha (M_Z^2)$ dominantly reflects, once 
again, the error in the data for $R(s)$.  The evaluation (\ref{eq:a9}) can be done for any $\sqrt{s_0} \gapproxeq 2$~GeV.  
Jergerlehner \cite{J} chooses $\sqrt{s_0} = 2.5$~GeV and finds
\bea
\label{eq:a10}
\Delta \alpha_{\rm had} (M_Z^2) & = & \Delta \alpha_{\rm had} (-M_Z^2) \: + \: (0.45 \pm 0.02) \times 10^{-4} \nonumber \\
& = & (277.82 \pm 2.54) \times 10^{-4},
\eea
where the error is entirely attributed to that on $\Delta \alpha_{\rm had} (- s_0)$.

Since the error in $\alpha (M_Z^2)$ is driven by the data for $R(s)$, it is relevant to discuss the new and forthcoming 
measurements.  First, the CND-2 and SND experiments at VEPP-2M at Novosibirsk have made precision measurements of a 
wide range of exclusive processes in the interval $0.4 < \sqrt{s} < 1.4$~GeV \cite{RHO,NOV}.  For example, the measurement of 
$e^+ e^- \rightarrow \pi^+ \pi^-$ \cite{RHO} has considerably improved precision such that, together with chiral 
perturbation theory, we find that the error in the $2 \pi$ contribution to $\Delta \alpha_{\rm had}$ for $\sqrt{s} 
< 0.96$~GeV is reduced to about $\pm 0.5 \times 10^{-4}$, see Table 1.

The second source of new information is the Beijing $e^+ e^-$ collider, where BES II are making direct measurements 
of $R(s)$ for $2 < \sqrt{s} < 5$~GeV.  So far values of $R(s)$ at six values of $\sqrt{s}$ have been published \cite{B98}.  
The two lowest values $(\sqrt{s} = 2.6, 3.2$~GeV) are shown in Fig.~1, together with preliminary BES II data \cite{B99}. 
\begin{figure}[h]
\label{fig:fig1}
\begin{center}
\mbox{\epsfig{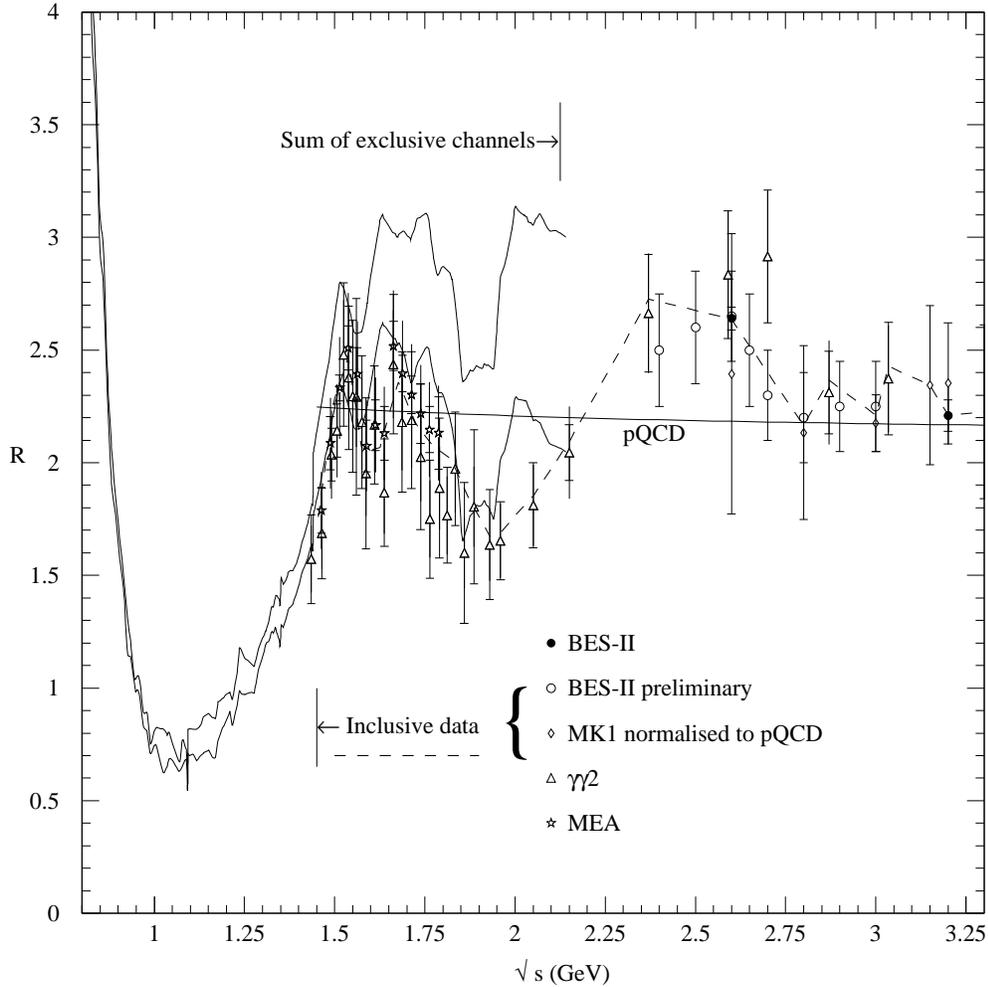}}
\caption{
A plot of the ratio $R(s)$ versus $\sqrt s$ in the most sensitive $\sqrt s$ interval.
Up to $\sqrt s=2.125$ GeV we show by continuous lines the upper and
lower bounds of the sum of exclusive channels. Above $\sqrt s=1.4$ GeV
we show the exclusive measurements of $R(s)$, together with the
interpolation (dashed curve) used in (4). We show six preliminary
BES-II points, which were not used in the analysis, together with
two published points which were used. The central perturbative QCD
prediction is also shown. In the region $1.46 < \sqrt s < 2.125$ GeV
we used, in turn, the inclusive and exclusive data to evaluate 
$\Delta \alpha_{\rm had}^{(5)}$.
}
\end{center}
\end{figure}
These data, taken together with the compatible earlier $\gamma\gamma 2$ data \cite{GG2}, indicate the existence of a 
\lq\lq dip-bump\rq\rq~type structure in the measurements of $R(s)$ about the ${\cal O} (\alpha_S^3)$ 
perturbative QCD prediction shown by the continuous curve, see Fig.~1.  The bump appears to be at $\sqrt{s} \sim 2.6$~GeV 
and the dip at $\sqrt{s} \sim 2$~GeV.  This apparent structure leads to an uncertainty in 
$\Delta \alpha_{\rm had}$.  Interestingly, the contributions to the right-hand-side of (\ref{eq:a4}) from the 
interval $1.8 < \sqrt{s} < 3$~GeV obtained using for $R(s)$ (i) the interpolation through the data and (ii) the 
perturbative QCD prediction are
\be
\label{eq:a12}
\Delta \alpha_{\rm had}^{1.8-3~{\rm GeV}} (M_Z^2) \; = \; 17.77 \pm 1.80 \quad {\rm and} \quad 
17.35 \pm 0.13,
\ee
respectively, in units of $10^{-4}$.  That is the results are very similar, which looks reasonable from the 
hadron-parton duality and QCD sum rule viewpoint.  We take the data value and so, in fact, the uncertainty from 
this domain may be less than our pessimistic estimate.

In the energy intervals described by perturbative QCD $(3 < \sqrt{s} < 3.74$~GeV and $\sqrt{s} > 5$~GeV) 
we use both the two-loop 
expression with the quark mass explicity included and the massless three-loop expression \cite{CHET} calculated in 
the $\overline{\rm MS}$ renormalization scheme\footnote{The uncertainty due to using a different scheme may be estimated 
to be of the order of the ${\cal O} (\alpha_S^4)$ correction, which is about $3 \sum e_q^2 r_3 (\alpha_S/\pi)^4$.  We may 
take $r_3 = -128$ \cite{KS2} which leads to an uncertainty much smaller than that given in Table 1.}.  In addition to the 
above we include the perturbative QCD error coming from varying $m_c, m_b, M_Z$ within the uncertainties quoted in the 
\cite{PDG}, $\alpha_S (M_Z^2) = 0.119 \pm 0.002$ and varying the scale $\alpha_S (cs)$ in the range $0.25 < c < 4$\footnote{We 
thank Thomas Teubner for valuable discussions concerning the perturbative QCD contribution.}. 

By far the largest uncertainty comes from the region $1.4 < \sqrt{s} < 3$~GeV.  The recent Novosibirsk measurements 
have considerably improved the exclusive data below $1.4$~GeV, but above $1.4$~GeV we see from Fig.~1 that the sum of 
the exclusive channels exceeds the inclusive measurements of $R$.  We therefore evaluated the contribution in the 
region $1.46 < \sqrt{s} < 2.125$~GeV using first the inclusive data\footnote{The inclusive measurements of the $\gamma \gamma 2$ collaboration 
(with radiative corrections included) at higher energies appear consistent with the new Beijing measurements, which 
would tend to favour the inclusive input leading to a smaller value of $\Delta \alpha_{\rm had}$.} for $R$ and then 
using the sum of the exclusive channels.  (We checked that our results for the contributions from the exclusive channels 
were in agreement with the detailed table of results given in Ref.~\cite{CH}, if we were to omit the new Novosibirsk data 
and to include the $\tau$ data.)

\begin{table} 
\caption{Contributions to $\Delta \alpha_{\rm had} (s) \times 10^4$ of (\ref{eq:a4})  
coming from the different $\sqrt{s^\prime}$ intervals for three different values of $s$. 
The alternative values in the round brackets take the summation of exclusive channels 
as the contribution from the region 1.46 - 2.125 GeV, rather than that from 
the inclusive data for $R(s^{\prime})$.} 
\begin{center} 
\begin{tabular}{|c|c|c|c|} \hline 
$\sqrt{s^\prime}$ interval & $s = M_Z^2$ & $s = -6$~GeV$^2$ & $s = -M_Z^2$ \\ 
(GeV) & & & \\ \hline 
$2m_{\pi} - 1.46^{a}$ & 38.18 $\pm$ $\left\{\begin{array}{c}0.52 \\ 0.60^{b} 
\end{array}\right\}$  
                  & 34.24 $\pm$ $\left\{\begin{array}{c}0.48 \\ 0.49^{b} \end{array}\right\}$ 
                  & 38.18 $\pm$ $\left\{\begin{array}{c}0.52 \\ 0.60^{b} \end{array}\right\}$\\ 
1.46 - 2.125 & 11.80 $\pm$ 0.94$^{c}$ & 7.80 $\pm$ 0.60$^{c}$ & 11.79 $\pm$ 
0.94$^{c}$ \\ 
& $\left(14.59 \pm 1.76^{b}\right) $ 
& $\left( 9.59 \pm 1.14^{b}\right) $ 
& $\left(14.58 \pm 1.76^{b}\right) $\\ 
2.125 - 3 & 13.05 $\pm$ 1.22$^{c}$ & 6.33 $\pm$ 0.58$^{c}$ & 13.03 $\pm$ 1.22$^{c}$ \\ 
3 - 3.74$^{d}$ & 7.41 $\pm$ 0.04 & 2.58 $\pm$ 0.01 & 7.39 $\pm$ 0.04 \\ 
3.74 - 5 &  15.11 $\pm$ 0.50 & 3.64 $\pm$ 0.12 & 15.04 $\pm$ 0.50 \\  
5 - $\infty^{d}$ & 170.06 $\pm$ 0.66 & 6.09 $\pm$ 0.03 & 169.83 $\pm$ 0.44 \\ 
$\omega$, $\phi$, $\psi$'s, $\Upsilon$'s & 18.57 $\pm$ 0.57 & 11.26 $\pm$ 0.27 
                                         & 18.52 $\pm$ 0.57 \\ \hline 
$\Delta \alpha_{\rm had}^{(5)} \times 10^4$ &  274.18 $\pm$ 2.52  
                                            &   71.94 $\pm$ 1.40
                                            &  273.78 $\pm$ 2.47 \\  
& $\left( 276.97 \pm  2.90 \right) $  
& $\left( 73.73  \pm  1.82 \right) $ 
& $\left( 276.57 \pm  2.85 \right) $\\ \hline  
\end{tabular} 
\end{center}
\scriptsize{$^{a}$ The upper (lower) error corresponds to the $2\pi$ (remaining) exclusive 
channels.} 

\scriptsize{$^{b,c,d}$ Errors with identical superscripts are added linearly. 
The remaining errors are added in quadrature.} 
\end{table}

The results for $\Delta \alpha_{\rm had} (s)$ of (\ref{eq:a4}) are shown in Table 1 not only for $s = M_Z^2$, but also 
for two space-like values $s \equiv - s_0 = -6$~GeV$^2$ and $s = -M_Z^2$ in order to study Jegerlehner's 
approach \cite{J}.  
We see that indeed the error on the space-like evaluation at $s = -6$~GeV$^2$ is reduced in comparison to that for 
$s = \pm M_Z^2$; as expected from the form of (\ref{eq:a4}) the error mainly arises from uncertainties in the data 
for $R (s^\prime)$ with $s^\prime \lapproxeq |s_0|$.  However before we can take advantage of this reduction we must 
consider the error in the perturbative continuation from $s = -s_0$ to $-M_Z^2$.  (It is reasonable to assume that 
the error in going round the circle to $s = M_Z^2$ is much smaller.)~~To estimate the error in the difference 
$\Delta \alpha_{\rm had} (-M_Z^2) - \Delta \alpha_{\rm had} (- s_0)$ we use the pure perturbative expression for 
$R (s^\prime)$ in the whole interval of $s^\prime$.  The resulting uncertainty is the order of $\pm 1 \times 
10^{-4}$ and so in practice there is only a marginal gain from the space-like evaluation.
Until this theoretical error is more accurately quantified, we take a 
conservative approach and use the error of the $s=M_Z^2$ determination
of $\Delta \alpha_{\rm had} (M_Z^2)$.

Let us return to the dip-bump structure in $R(s)$ in the region $2 < \sqrt{s} < 3$~GeV, see Fig.~1.  This was not 
expected, at least from the quark states; 
recall the success of the perturbative QCD description of $\tau$ hadronic decay.  Charm is too heavy, and the light 
quarks too light to produce such a {\it broad} structure.  Perturbative QCD was expected to reproduce the $R(s)$ data 
in this $\sqrt{s}$ interval.  However at ${\cal O} (\alpha_S^3)$ we open up the 3-gluon contribution, see Fig.~2, and we 
may expect some structure due to this channel.  Indeed lattice computations predict a $J^P = 1^-$ glueball with mass 
typically in the range 3.3--4~GeV \cite{LAT}.  So non-perturbative gluonic structure at ${\cal O} (\alpha_S^3)$, and 
higher order, may be anticipated in the region up to 4~GeV.  In this connection we note that the non-perturbative 
(condensate) contribution to the Adler function $D (-s)$ diverges for values of $\sqrt{s} \lapproxeq 2$~GeV, even in 
the space-like region \cite{EJKV}.
\begin{figure}[h]
\label{fig:fig2}
\begin{center}
\mbox{\epsfig{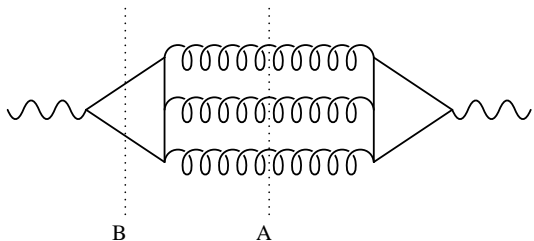}}
\caption{
${\cal O} (\alpha_S^3)$ three gluon contribution to $R$.  The contribution of the light quarks 
cancel as $\sum e_q = 0$ for $q = u, d, s$, so we are left with the $c$ quark contribution.
}
\end{center}
\end{figure}
Recall in the phenomenological description of the decay $J/\psi \rightarrow gg \gamma$ that it was crucial to introduce 
an effective gluon mass of about 800~MeV to reproduce the $\gamma$ energy spectrum \cite{ggG}.  A similar result 
holds for $\Upsilon$ and $Z$ decays \cite{UP,ZD}.  We thus anticipate an effective threshold for the 3 gluon 
contribution to $R(s)$ at $\sqrt{s} \simeq 2.4$~GeV.  This effective gluon mass is also consistent with the 
predicted 1.6~GeV mass of the lightest ($0^{++}$) glueball.  More or less the same position of the threshold in 
$R(s)$ is obtained assuming that the hadronization of the 3 gluon contribution to $R$ (cut $A$ in Fig.~2) proceeds 
through the 6-quark channel.  In the threshold region we must use the constituent quark mass, so that the 3 gluon 
channel opens at $\sqrt{s} \gapproxeq 6 M_{\rm constit} \simeq 2.1$~GeV.

Fig.~2 has not only the possibility of a gluon cut, but instead may be cut across the $q\bar{q}$ state, cut $B$.  
With respect to this latter cut, the imaginary part in the gluon channel plays the role of an absorptive correction.  
Therefore the diagram may give a negative $q\bar{q}$ contribution and a positive $3g$ contribution leading to a 
non-trivial structure\footnote{Recall the cusp structure for a resonance occurring near a threshold in multichannel 
processes.}.  In addition there may be mixing, and mass shifts, of the $J^P = 1^{- -}$ glueball and $q\bar{q}$ 
resonant states.  Of course the amplitude of this gluonic-driven structure cannot be too large as it originates 
from $\alpha_S^3$ and higher order graphs.  However the non-perturbative 3 gluon interactions are strong enough 
to produce a resonant type structure in the relevant $\sqrt{s}$ region with up to about 10\%\footnote{Estimated from 
the difference between ${\cal O} (\alpha_S^3)$ and ${\cal O} (\alpha_S^2)$ QCD expectations.} oscillations, 
and seen in Fig.~1.  Note that the non-perturbative contribution to the Addler $D$ function becomes large at 
$\sqrt{|s|} \sim 2$~GeV even in the space-like region \cite{EJKV}.  If the bump is due to 3-gluon production then we 
expect a larger yield of $\eta^\prime$ or $\eta$ mesons (which are strongly coupled to gluons) in this energy 
region \cite{ETA}.

We conclude that it is important to use the experimental measurements of $R(s)$ to calculate $\Delta \alpha_{\rm had}$ 
and its error.  One can replace the data for $R(s)$ in a specific region by the QCD prediction only after we understand 
the origin of all structure in the region, and are confident that our QCD approximation is sufficient to account for 
the effects.  For example, much accuracy is gained by replacing the data measurements for $R(s)$ above the $\Upsilon$ 
resonance region by the ${\cal O} (\alpha_S^3)$ QCD prediction.  However we have seen that a similar replacement around the 
charm threshold is not so clear.  The theoretical uncertainty due to the charm mass and the (more than two-loop) 
non-perturbative interaction may be larger than the accuracy of the data.  From the presently available data we find 
(see Table 1)
\be
\label{eq:a16}
\Delta \alpha_{\rm had}^{(5)} (M_Z^2) \; = \; 274.18 \; \pm \; 2.52 \quad {\rm or} \quad 276.97 \; \pm \; 2.90
\ee
according to whether the inclusive or exclusive data are used in the interval $1.46 < \sqrt{s} < 2.125$~GeV.  When 
used in (\ref{eq:a2}) these values correspond to a value of the QED coupling at the $Z$ pole of
\be
\label{eq:a17}
\alpha (M_Z^2)^{-1} \; = \; 128.973 \; \pm \; 0.035 \quad {\rm or} \quad 128.934 \; \pm \; 0.040
\ee
respectively.  It may be argued that the first value is favoured, since the $\gamma \gamma 2$ inclusive data 
appear to agree with the new Beijing measurements at higher energies, see Fig.~1. \\

\noindent {\large \bf Acknowledgements}

We thank Andrei Kataev, Thomas Teubner and Zhengguo Zhao for valuable information.  One of us (MGR) thanks the Royal Society for 
support. \\
\noindent {\large \bf References} \\

\newpage

\end{document}